# Lightweight Vulnerability Detection from Code Metrics and Token Features

Chun Yin Chiu

## Abstract

Vulnerability detection for C/C++ code increasingly relies on heavy representations such as code graphs and deep models, while many practical workflows still benefit from fast and reproducible ranking baselines for human triage. This preprint studies a lightweight function-level vulnerability triage pipeline that combines sparse token n-grams from raw function text with a small set of inexpensive code metrics, including NLOC, approximate cyclomatic complexity, token count, maximum brace depth, and parameter count. We use TF-IDF token features and a class-weighted logistic regression classifier, avoiding deep learning, transformers, and program graphs.

Using the Devign function-level labels, we evaluate random and cross-project settings, including a FFmpeg-to-QEMU transfer experiment. We emphasize precision-recall AUC and Recall@10% as ranking-oriented metrics for skewed or triage-oriented workloads. On the random split, the best combined variant reaches PR-AUC 0.642 and Recall@10% 0.161, while cross-project generalization is substantially harder, with PR-AUC around 0.436. We further report ablations, test-only identifier-renaming robustness, and end-to-end efficiency. The results suggest that simple token and metric features provide a useful transparent baseline, but also expose sensitivity to superficial lexical cues and limited cross-project transfer.

## 1. Introduction

Software vulnerability detection has progressed rapidly with learning-based approaches that ingest rich program representations such as code property graphs and graph neural networks (GNNs), including Devign [1]. However, graph construction and deep training can be expensive, and the resulting systems are harder to deploy in resource-constrained environments or continuous integration pipelines. In contrast, many organizations primarily need a fast triage signal that ranks potentially vulnerable functions for human review, often under domain shift and label noise. This motivates lightweight detectors built from inexpensive features and classical models that are easy to reproduce and audit.

This work revisits vulnerability detection using only code metrics and token features, deliberately avoiding deep learning, code graphs, and heavyweight parsing. In contrast to graph neural approaches that learn on code-property graphs, we study a sparse, linear classifier trained on TF-IDF token n-grams and a handful of cheap function-level metrics.

Although the positive rate in our provided splits is moderate, real-world vulnerability triage often exhibits heavy imbalance. Following best practice for skewed settings, we emphasize precision-recall analysis and report PR-AUC as the primary metric [2], along with thresholded measures such as F1 and Recall@10%. Prior studies on vulnerability prediction with metrics and lexical features report substantial dataset shift and project-specific effects [3,4]. We also draw on established relationships between precision-recall and ROC evaluation [5] and common considerations in imbalanced learning [6].

This preprint is intended as a transparent baseline and diagnostic study rather than a claim of state-of-the-art vulnerability detection performance. Its contributions are:

- A lightweight vulnerability triage pipeline that combines TF-IDF token uni/bi-grams and a small set of cheap code metrics in a class-weighted logistic regression model.
- A focused empirical study covering ablation, cross-project transfer, test-only identifier renaming, and efficiency under a fixed Devign-based setup.
- A discussion of practical trade-offs, including lexical sensitivity, comment-retention risks, domain shift, and the limits of using simple features as standalone detectors.

## 2. Background and Related Work

Vulnerability prediction has long investigated whether simple signals such as complexity, churn, and developer activity can forecast security defects, often observing strong project dependence [3,10-13]. Text mining on code tokens provides another lightweight signal and has been shown effective at the component level [4,18]. Traditional static analyzers such as ITS4 [8] and Flawfinder [9] are highly interpretable but context-limited [7]. Recent detectors leverage deep learning and richer program representations, including Devign, VulDeePecker, and SySeVR [1,14,15], while clone-based approaches such as VUDDY target vulnerability propagation [16]. We position our method as a transparent, inexpensive baseline for ablation, dataset shift, and robustness studies, complementing audit-prioritization methods such as VCCFinder and exploitability-oriented analyses [17,19].

## 3. Proposed Method

Pipeline overview. Each function is processed by two linear-time feature extractors: (i) TF-IDF token n-grams over a regex token stream and (ii) five scalar code metrics. The vectors are concatenated and classified with class-weighted logistic regression, avoiding parsing-heavy representations such as ASTs, CFGs, CPGs, and any deep learning.

### 3.1. Token Features (TF-IDF)

We compute token features directly from the raw function text and intentionally do not remove comments. String literals are not treated as single atomic tokens; instead, they are tokenized by the same lexical regex without special normalization. Numeric literals are retained as tokens. We preserve case, so Foo and foo are distinct. The tokenizer is a simple regex-based lexer that extracts identifiers, numeric constants, common multi-character operators, and remaining punctuation. We evaluate unigram and unigram+bigram settings. N-grams are formed over the linear token stream, and bigrams may cross newline boundaries. Term importance is captured with TF-IDF weighting in the classic information-retrieval formulation [20].

Because comments are retained in the token branch, token features may capture natural-language hints or dataset-specific annotations rather than executable code alone. We therefore treat the token-based variants as lexical baselines and discuss comment retention as a construct-validity threat. A comment-stripped token setting is a necessary follow-up experiment.

### 3.2. Code Metric Features

We compute five inexpensive metrics per function: number of non-blank lines of code (NLOC), an approximate cyclomatic complexity score, token count, maximum brace depth, and parameter count. Unlike token features, metric extraction first strips both // and /* ... */ comments, so metrics reflect uncommented code structure. All metrics are computed with linear scans and simple counters, without building syntax trees.

### 3.3. Classifier

We use logistic regression with L2 regularization and class weighting (class_weight='balanced') to mitigate label skew. For token features we use scikit-learn's TfidfVectorizer with the custom tokenizer (token_pattern=None, lowercase=False, min_df=2). We use (1,1) or (1,2) n-grams for the respective variants. For combined models, numeric metrics are scaled using MaxAbsScaler and concatenated with the sparse TF-IDF vector. The classifier uses solver='liblinear', C=1.0, max_iter=2000, n_jobs=1, and random_state=42.

At test time, we rank functions by predicted vulnerability probability and evaluate Recall@10% for triage-oriented inspection budgets.

## 4. Experimental Setup

Dataset. We use the Devign benchmark with function-level vulnerability labels [1]. The provided results cover two evaluation settings: a random stratified split over the available functions, and a cross-project split training on FFmpeg and testing on QEMU. For the random split, the training set contains 21,854 functions with 9,968 vulnerable functions (45.6% positive), and the test set contains 2,732 functions with 1,246 vulnerable functions (45.6% positive). For the cross-project setting (FFmpeg-to-QEMU), the training set contains 9,769 functions with 4,981 positives (51.0%), and the test set contains 17,549 functions with 7,479 positives (42.6%).

Splits. For the random split we perform a stratified 80/10/10 split (train/validation/test) with a fixed seed (SEED=42) using train_test_split(..., stratify=y); the reported metrics are computed on the final test partition. For cross-project evaluation, we train on FFmpeg and evaluate on QEMU without mixing projects.

Variants. We evaluate four lightweight variants: metrics-only, tokens-only with unigrams, tokens-only with uni+bi-grams, and a combined model using metrics plus token uni+bi-grams. All variants use the same logistic regression classifier and class weighting.

Metrics. We report PR-AUC as the primary summary metric, alongside F1 and Recall@10%, which measures recall among the top 10% highest-scoring functions. This reflects a fixed inspection budget. We also include ROC-AUC and precision/recall at the default decision threshold in our local logs for completeness, but focus the paper on ranking-oriented measures. We adopt PR-AUC because it is more informative than ROC-AUC under skew [2,5], and Recall@10% reflects realistic triage budgets; for broader discussion of imbalanced learning challenges see [6].

Robustness. To assess sensitivity to superficial naming changes, we apply a lexical perturbation on the test set only: every token matching an identifier pattern that is not in a keyword list is replaced with a single placeholder token ID. This is implemented via direct regex replacement without parsing; consequently, identifiers appearing in comments or string literals may also be replaced.

Implementation and environment. TF-IDF uses min_df=2 with default scikit-learn settings otherwise. Training time includes TF-IDF fitting and logistic regression optimization. Inference time includes TF-IDF transformation and prediction on the test set. Feature extraction time measures only the computation of code metrics after comment stripping, computed once for the entire dataset prior to splitting; it excludes dataset I/O and model fitting. Experiments were run on a MacBookPro16,2 (x86_64) with an Intel Core i5-1038NG7 CPU, 32 GB RAM, and macOS 26.2, using Python 3.13.0, scikit-learn 1.8.0, NumPy 2.4.2, and pandas 3.0.0. SEED=42 is fixed for all randomized operations.

## 5. Empirical Study and Results

The goal of this empirical study is not to claim state-of-the-art vulnerability detection performance, but to characterize how far simple token and metric features can go under random, cross-project, robustness, and efficiency settings.

### 5.1. Main Performance and Ablation

Table 1 reports performance for the random and cross-project settings on original code. On the random split, combining metrics with token uni+bi-grams yields the best PR-AUC (0.642) and Recall@10% (0.161). The gain over metrics-only in PR-AUC is modest (0.005), suggesting that cheap metrics already capture much of the in-distribution signal. In the cross-project setting, performance drops substantially for all variants (PR-AUC approximately 0.42-0.44), indicating strong domain shift. The combined model provides the best PR-AUC (0.436), but differences are small, and Recall@10% remains close across variants.

Although the combined model provides the best random-split ranking score, Recall@10% remains low in absolute terms. This suggests that the pipeline is better viewed as a transparent baseline or auxiliary triage signal rather than a standalone vulnerability detector.

| Variant | Random PR-AUC | Random F1 | Random R@10% | Cross PR-AUC | Cross F1 | Cross R@10% |
|---|---|---|---|---|---|---|
| Metrics | 0.638 | 0.616 | 0.156 | 0.434 | 0.418 | 0.105 |
| Tok-U | 0.634 | 0.611 | 0.154 | 0.423 | 0.436 | 0.098 |
| Tok-UB | 0.638 | 0.617 | 0.158 | 0.423 | 0.418 | 0.096 |
| Mix | 0.642 | 0.619 | 0.161 | 0.436 | 0.398 | 0.105 |

### 5.2. Cross-Project Generalization

Table 2 details the FFmpeg-to-QEMU setting. Compared to the random split, PR-AUC drops from approximately 0.64 to approximately 0.43-0.44 across variants, highlighting the difficulty of transferring vulnerability signals across codebases. The best PR-AUC in this setting is achieved by the combined model (0.436), but the margin over metrics-only is negligible. This result is consistent with prior observations that vulnerability-prediction features can be highly project-dependent.

| Train project | Test project | Variant | PR-AUC | F1 | R@10% |
|---|---|---|---|---|---|
| FFmpeg | QEMU | Metrics | 0.434 | 0.418 | 0.105 |
| FFmpeg | QEMU | Tok-U | 0.423 | 0.436 | 0.098 |
| FFmpeg | QEMU | Tok-UB | 0.423 | 0.418 | 0.096 |
| FFmpeg | QEMU | Mix | 0.436 | 0.398 | 0.105 |

### 5.3. Robustness to Identifier Renaming

Table 3 reports the effect of test-only identifier renaming. In the random split, all variants lose PR-AUC and Recall@10%, showing that lexical signals tied to specific names contribute to ranking. In cross-project evaluation, renaming slightly increases PR-AUC for all variants, suggesting that removing project-specific identifiers can reduce spurious correlations when transferring across codebases. Notably, metrics-only yields the best PR-AUC under renaming in both splits.

We also observe that threshold-dependent metrics such as F1 at a fixed cutoff can change sharply under renaming and cross-project shift, even when ranking quality degrades more smoothly. This suggests that lightweight detectors intended for triage should be evaluated primarily as rankers and, if used for binary decisions, should incorporate probability calibration or threshold selection on a held-out validation set from the target domain.

| Setting | Variant | PR-AUC orig | PR-AUC rename | Delta | R@10% orig | R@10% rename | Delta |
|---|---|---|---|---|---|---|---|
| Random | Metrics | 0.638 | 0.550 | -0.088 | 0.156 | 0.139 | -0.017 |
| Random | Tok-U | 0.634 | 0.525 | -0.108 | 0.154 | 0.127 | -0.027 |
| Random | Tok-UB | 0.638 | 0.527 | -0.111 | 0.158 | 0.130 | -0.028 |
| Random | Mix | 0.642 | 0.549 | -0.093 | 0.161 | 0.135 | -0.026 |
| Cross | Metrics | 0.434 | 0.461 | 0.028 | 0.105 | 0.113 | 0.008 |
| Cross | Tok-U | 0.423 | 0.444 | 0.021 | 0.098 | 0.108 | 0.010 |
| Cross | Tok-UB | 0.423 | 0.433 | 0.010 | 0.096 | 0.103 | 0.006 |
| Cross | Mix | 0.436 | 0.458 | 0.023 | 0.105 | 0.112 | 0.007 |

### 5.4. Efficiency Analysis

Table 4 summarizes runtime costs. Metric extraction takes 17.52 s for the full dataset and is shared across variants. Tokens with bi-grams roughly double training and inference time compared to unigrams because the feature space expands. Even so, end-to-end training remains in the tens of seconds on a laptop CPU, and inference is below 2.5 s for the random-split test set. Because the feature extractors are linear in input size and the classifier is convex, runtime scales predictably with dataset size and vocabulary. In practice, this makes the approach suitable for nightly scans or pull-request gating where thousands of functions must be scored quickly.

| Setting | Variant | Feat. time (s) | Train time (s) | Infer. time (s) | n_train/n_test |
|---|---|---|---|---|---|
| Cross | Metrics | 17.52 | 6.67 | 6.26 | 9769/17549 |
| Cross | Mix | 17.52 | 11.88 | 9.88 | 9769/17549 |
| Cross | Tok-U | 17.52 | 6.62 | 6.18 | 9769/17549 |
| Cross | Tok-UB | 17.52 | 11.95 | 9.81 | 9769/17549 |
| Random | Metrics | 17.52 | 12.69 | 1.57 | 21854/2732 |
| Random | Mix | 17.52 | 23.62 | 2.16 | 21854/2732 |
| Random | Tok-U | 17.52 | 12.57 | 1.62 | 21854/2732 |
| Random | Tok-UB | 17.52 | 22.60 | 2.46 | 21854/2732 |

The previous draft included precision-recall and efficiency figures. In this cleaned preprint version, the inconsistent precision-recall plot has been removed to avoid mixing curve-integration AUC with average precision. The tables above consistently use average precision as PR-AUC.

## 6. Threats to Validity

Internal validity. We report results exactly as recorded in the provided CSV. Timing measurements can vary with background load, caching, and implementation details; we partially mitigate this by reporting per-variant times under a fixed environment and by describing what each timer includes and excludes.

Construct validity. Ground-truth labels inherit noise from the underlying benchmark construction and may not cover all vulnerability types. F1 depends on a fixed decision threshold and may be brittle under domain shift; Recall@10% is more aligned with triage but may not reflect all deployment policies. Comment retention may introduce non-code cues, especially if comments contain bug, fix, security, or vulnerability-related language. This may inflate in-distribution performance and partly explain the sensitivity of token features to identifier renaming and cross-project shift.

External validity. Our evaluation is limited to the projects present in the provided results: a random split and FFmpeg-to-QEMU transfer. Other languages, repositories, and vulnerability taxonomies may exhibit different feature-label relationships. Identifier renaming is an artificial perturbation and may not fully capture realistic refactoring.

## 7. Conclusion and Future Work

We presented a lightweight vulnerability detector built from TF-IDF token features and inexpensive code metrics, trained with class-weighted logistic regression. Across the provided experiments, the combined model performs best on the random split (PR-AUC 0.642, Recall@10% 0.161), while cross-project generalization remains challenging (PR-AUC approximately 0.43-0.44). Robustness experiments show that token features are sensitive to identifier renaming, whereas metrics provide more stable signals.

Overall, the results support the use of simple token and metric features as transparent vulnerability-triage baselines, not as standalone detectors. Future work will (i) add a comment-stripped token setting to test potential leakage from comments, (ii) normalize numbers and string literals with NUM/STR placeholders to reduce overfitting to surface forms, (iii) extend evaluation to additional projects and real-world imbalanced settings, and (iv) evaluate calibrated decision thresholds on held-out target-domain validation data.